\documentclass[fleqn,usenatbib]{mnras}
\usepackage{newtxtext,newtxmath}
\usepackage[T1]{fontenc}
\usepackage{gensymb}

\DeclareRobustCommand{\VAN}[3]{#2}
\let\VANthebibliography\thebibliography
\def\thebibliography{\DeclareRobustCommand{\VAN}[3]{##3}\VANthebibliography}

\usepackage{CJKutf8}
\usepackage{booktabs}
\usepackage{xcolor, soul}
\usepackage{amsmath, bm}
\usepackage{graphicx}

\title[Cosmological constraints from weak lensing scattering transform]
{Cosmological constraints from weak lensing scattering transform using HSC Y1 data}

\author[Cheng et al.]{
Sihao Cheng (程思浩)$^{1,2}$\thanks{E-mail: scheng@ias.edu}, Gabriela A. Marques$^{3,4}$, Daniela Grand\'on$^{5,6}$, Leander Thiele$^{7,8}$, 
\newauthor
Masato Shirasaki$^{9,10}$, Brice M\'enard$^{11}$, \& Jia Liu$^{8}$
\\
$^{1}$Institute for Advanced Study, 1 Einstein Dr., Princeton, NJ 08540, USA\\
$^{2}$Perimeter Institute for Theoretical Physics, 31 Caroline St N, Waterloo, ON N2L 2Y5, Canada\\
$^{3}$Fermi National Accelerator Laboratory, P. O. Box 500, Batavia, IL 60510, USA\\
$^{4}$Kavli Institute for Cosmological Physics, University of Chicago, Chicago, IL 60637, USA\\
$^{5}$Mathematical Institute, Leiden University, Gorlaeus Gebouw, Einsteinweg 55, NL-2333 CC Leiden, The Netherlands\\
$^{6}$Departamento de F\'isica, FCFM, Universidad de Chile, Blanco Encalada 2008, Santiago, Chile\\
$^{7}$Department of Physics, Princeton University, Princeton, NJ 08544, USA\\
$^{8}$Center for Data-Driven Discovery, Kavli IPMU (WPI), UTIAS, The University of Tokyo, Kashiwa, Chiba 277-8583, Japan\\
$^{9}$National Astronomical Observatory of Japan (NAOJ), National Institutes of Natural Sciences, Osawa, Mitaka, Tokyo 181-8588, Japan\\
$^{10}$The Institute of Statistical Mathematics, Tachikawa, Tokyo 190-8562, Japan\\
$^{11}$Department of Physics and Astronomy, The Johns Hopkins University, 3400 N. Charles St., Baltimore, MD 21218, USA
}

\pubyear{2024}

\begin{document}
\label{firstpage}
\pagerange{\pageref{firstpage}--\pageref{lastpage}}
\begin{CJK}{UTF8}{gkai} 
\maketitle
\end{CJK}

\begin{abstract}
As weak lensing surveys go deeper, there is an increasing need for reliable characterization of non-Gaussian structures at small angular scales. Here we present the first cosmological constraints with weak lensing scattering transform, a statistical estimator that combines efficiency, robustness, and interpretability. With the Hyper Suprime-Cam survey (HSC) year 1 data, we obtain $\Omega_\text{m}=0.29_{-0.03}^{+0.04}$, $S_8\equiv \sigma_8(\Omega_\text{m}/0.3)^{0.5}=0.83\pm0.02$, and intrinsic alignment strength $A_\text{IA}=1.0\pm0.4$ through simulation-based forward modeling. 
Our constraints are consistent with those derived from \textit{Planck}. 
The error bar of $\Omega_\text{m}$ is 2 times tighter than that obtained from the power spectrum when the same scale range is used. This constraining power is on par with that of convolutional neural networks, suggesting that further investment in spatial information extraction may not yield substantial benefits.

We also point out an internal tension of $S_8$ estimates linked to a redshift bin around $z\sim1$ in the HSC data. We found that discarding that bin leads to a consistent decrease of $S_8$ from 0.83 to 0.79, for all statistical estimators.
We argue that 
photometric redshift estimation is now the main limitation in the estimation of $S_8$ using HSC. This limitation is likely to affect other ground-based weak lensing surveys reaching redshifts greater than one. Alternative redshift estimation techniques, like clustering redshifts, may help alleviate this limitation.

\end{abstract}


\section{Introduction} 
\label{sec:intro}

\begin{figure*}
    \centering
    \includegraphics[width=\textwidth]{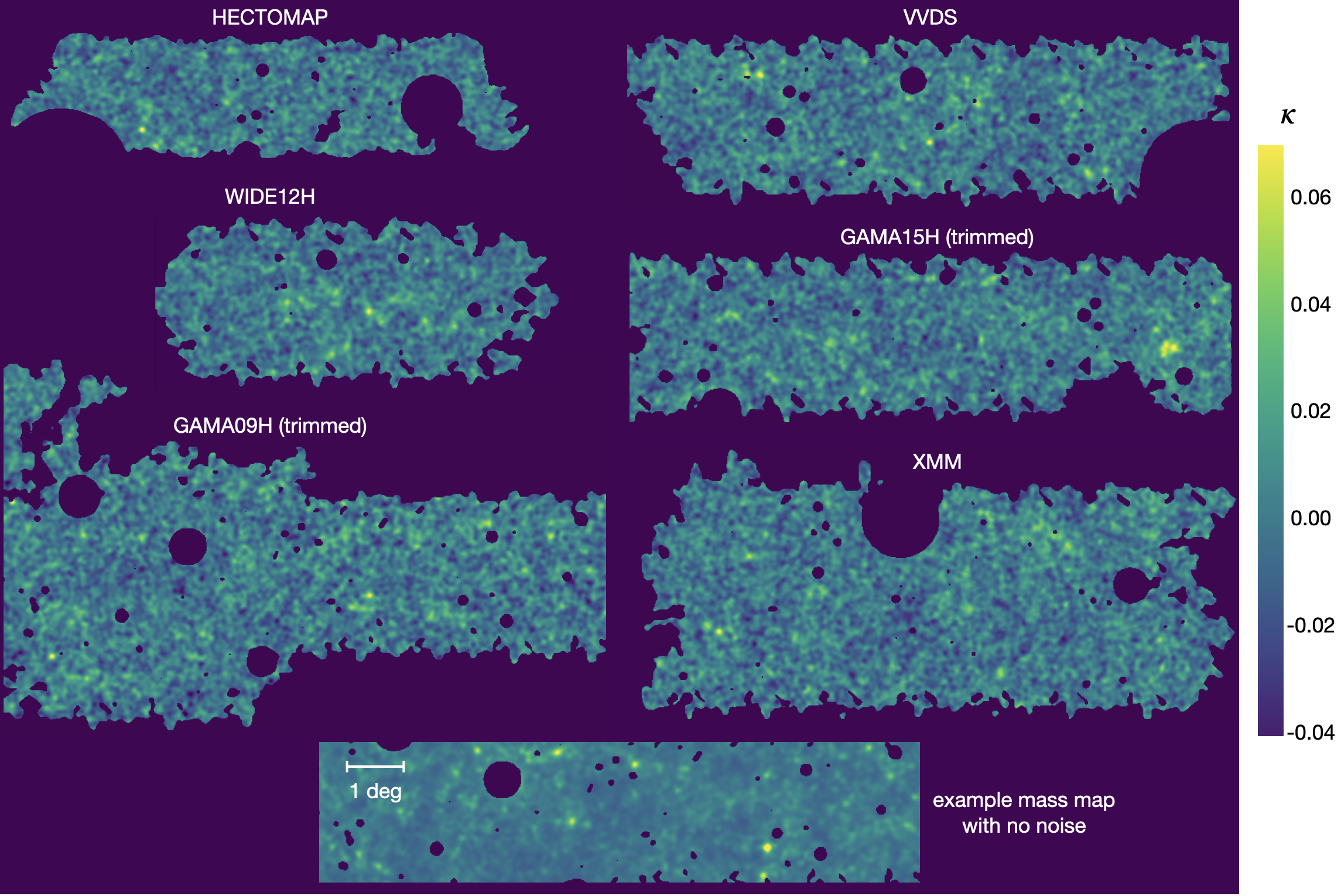}
\caption{The convergence maps made from HSC Y1 galaxy shear catalog (six panels), and a part of example mock map tailored to GAMA09H with no galaxy shape noise for reference.  To visually demonstrate the cosmic shear signal, we only show maps using galaxies in the full photometric redshift range of 0.3--1.5 and thus with the highest signal-to-noise ratio. The maps are made using the procedure described in section~\ref{sec:map} with 2 arcmin Gaussian smoothing. There are visible non-Gaussian structures, such as hierarchically clustered peaks above the noise level.}
    \label{fig:maps}
\end{figure*}

Due to gravitational lensing, matter density fluctuations in our universe distort the images of background galaxies. On cosmological scales, this phenomenon is referred to as cosmic shear \citep[for a review, see][]{kilbinger2015cosmology,review_wlmandelbaum}. This effect can be used to probe the geometry and the matter distribution of our Universe. At relatively large angular scales, the lensing field is approximately Gaussian and can be characterized by two-point statistics, such as the power spectrum or correlation function. However, as illustrated in Figure~\ref{fig:maps}, ongoing weak lensing surveys can probe scales small enough to reveal rich non-Gaussian structures, requiring additional statistics to extract the full cosmological information in the field.

As opposed to Gaussian fields, non-Gaussianities possess much more degrees of freedom, and different statistics may exhibit sensitivity to diverse features and systematic effects in a map.
The efficacy of non-Gaussian statistics has been explored in the weak lensing context, including on the most recent datasets
\citep{martinet2018kids,fluri2022full,zurcher2022dark,secco2022_DES3pt,liu2023cosmological,Thiele_2023,anbajagane2023beyond,heydenreich2022persistent,burger2023kids,Marques_2024}. However, there has been no simple agreement on which summary statistic to use. Traditional moment based statistics are not robust to outliers in real data. Other simple statistics are usually too focused on particular features, and thus cannot extract the full cosmological information in a field. On the other extreme, inference with deep learning methods provide better constraining power but still lack interpretability.

Recently, a new summary statistic called the scattering transform has been proposed to extract cosmological information from weak lensing maps \citep{Cheng_2020}.  Similar statistics using related ideas, such as the phase harmonics \citep{Allys_2020} and scattering spectra \citep{Cheng_2024}, have also been proposed and studied on mock data for cosmology.  Originating from the computer vision community \citep{Bruna_2013}, the scattering transform is shown to exhibit high information content and robustness in mock cosmological \citep{Cheng_2021, Gatti_2023st} and astrophysical \citep{Allys_2019} analysis. 
In this paper, we present the first application of this statistic to observational weak lensing data\footnote{Application of the scattering transform has been explored to galaxy clustering with BOSS data \citep{valogiannis2022going, Valogiannis_2023} but not yet on weak lensing data.}. Our analysis uses the Subaru Hyper Suprime-Cam first-year (HSC Y1) data. We employ a forward modeling approach with a large set of $N$-body simulations and mock catalogs customized for the data, and derive cosmological constraints.
\\
\\
Weak lensing data is mainly sensitive to two cosmological parameters and provides roughly uncorrelated constraints on $S_8\equiv \sigma_8\sqrt{\Omega_m/0.3}$ and $\Omega_\text{m}$. Here, $\Omega_m$ is the total matter density today, and $\sigma_8$ represents the linear matter fluctuation within a sphere of radius $8 h^{-1}$Mpc. The scattering transform
and in general non-Gaussian statistics are expected to significantly tighten the constraint of $\Omega_\text{m}$ by distinguishing fields with similar amplitude of lensing fluctuations but different extent into the non-linear regime. In addition, in light of the recent tension between $S_8$ values derived from the cosmic microwave background \citep{Planck_2020} and large-scale structures found in various studies \citep[e.g.,][]{hildebrandt2020kids+,asgari2021kids,garcia2021growth,des3x2, amon2022dark,Li_2023,Dalal_2023, marques2024cosmological}, the exploration of non-Gaussian statistics provides a consistency check and may open a new avenue for understanding the underlying cause.



\section{statistical analysis of HSC Y1 data} \label{sec:data}

We perform a statistical analysis of the first year data of the HSC survey\footnote{\url{https://hsc.mtk.nao.ac.jp/ssp/}} \citep{Aihara_2018a}, which is a wide-area imaging survey conducted using the 8.2m Subaru telescope in Hawaii in five broad-bands, $grizy$. The HSC is the the deepest large-area weak lensing survey to date, with 5$\sigma$ point-source depth of $i\sim$ 26 and typically $i$-band seeing of 0.58". Hence, the HSC data is well suited for studying the non-Gaussian structures in the lensing field.

The first year data release of HSC, referred to as HSC S16A \citep{Mandelbaum_2018HSC}, contains data taken from March 2014 to April 2016 and covers an area of 136.9 deg$^2$ area in six disjoint fields: \texttt{GAMA09H}, \texttt{GAMA15H}, \texttt{HECTOMAP}, \texttt{VVDS}, \texttt{WIDE12H}, and \texttt{XMM}. Due to the limited 10$\times$10 deg$^2$ size of our cosmo-varied simulations, we trim off some edge regions in \texttt{GAMA09H} and \texttt{GAMA15H} that exceed a width of 10 degrees.

\subsection{Weak lensing mass map}
\label{sec:map}

Our analysis is based on weak lensing convergence maps (also called mass maps) created using the galaxy shapes measured by \citet{Mandelbaum_2018HSC} from the $i$-band coadded images with the re-Gaussianization PSF correction method \citep{Hirata_2003}. Figure~\ref{fig:maps} shows the resulting mass maps, which clearly reveals non-Gaussian structures such as peaks and ridges. Below, we describe the procedures for creating the mass maps. 

To perform a tomographic lensing analysis, we select source galaxies in 4 redshift bins, with edges [0.3, 0.6, 0.9, 1.2, 1.5]. 
Below, we will refer to those bins as bin 1 to 4, from low to high redshift.
The redshift bin assignment is determined using the ``best-fit'' point estimate of redshift provided by the \texttt{Ephor-AB} photo-z code \citep{Tanaka_2018}. The effective number density of galaxies in each photo-z bin is 5.14, 5.23, 3.99, 2.33 $\text{arcmin}^{-2}$, respectively. The total number of galaxies is about 8.5 million.

We begin by creating shear maps from the galaxy shape catalog. The survey area is discretized into pixels, each with a size of 0.88 arcminutes. Then, we calibrate and average 
the observed galaxy ellipticity $\boldsymbol{e}$ to obtain an unbiased estimator $\hat{\boldsymbol{\gamma}}$ of the underlying cosmic shear $\boldsymbol{\gamma}$ for each pixel,
\begin{align}
    \hat{\boldsymbol{\gamma}} = \frac{1}{1+ \bar{m}_\text{tot}}\bigg(\frac{\bar{\boldsymbol{e}}}{2 \bar{\mathcal{R}}} -\bar{\boldsymbol{c}}\bigg)\,,
    \label{eq:map_pixel}
\end{align}
where $1+m_\text{tot}$ is the total multiplicative bias, $\boldsymbol{c}$ is the additive bias for each galaxy, and $\mathcal{R}\equiv 1-e_\text{rms}^2$ denotes the responsivity. This estimator originates from taking expectation of both sides of the shear distortion equation at small shear limit: $\boldsymbol{e} = \boldsymbol{e}^\text{int} + 2(1+m)\mathcal{R}\boldsymbol{\gamma} + \boldsymbol{c} + \boldsymbol{e}^\text{mea}$, where $\boldsymbol{e}^\text{int}$ and $\boldsymbol{e}^\text{mea}$ are the intrinsic galaxy ellipticity and measurement error, both assumed to have zero expectation.

Following \citet{Hikage_2019}, we include three terms in the total multiplicative bias $m_\text{tot}= m + m_\text{sel} + m_\mathcal{R}$, where $m$ is that obtained from image simulations \citep{Mandelbaum_2018b} and averaged within each of the individual fields and for each of the four redshift bins, $m_\text{sel}=[0.0086, 0.0099, 0.0091, 0.0091]$ is the multiplicative bias caused by galaxy size selection for each redshift bin, and $m_\mathcal{R}=[0.000, 0.000, 0.015, 0.030]$ is the correction for redshift-dependent responsivity \citep[see section 5.3 of][]{Mandelbaum_2018b}.

In eq.~\ref{eq:map_pixel}, the bar $\bar{\cdot}$ represents averages weighted by $w$:
\begin{align}
    \bar{\phi}(x,y) \equiv \frac{\sum_\text{in pixel (x,y)} w_i \phi_i}{\sum_\text{in pixel (x,y)} w_i}\,,
    \label{eq:weight}
\end{align}
where $\phi(x,y)$ represent the value of any quantity in $\{\boldsymbol{e}, \mathcal{R}, m_\text{tot}, \boldsymbol{c} \}$ at the pixel (x,y), and the weights $w_i$ are read directly from the HSC S16A catalog \citep{Mandelbaum_2018HSC}, where it is defined as $w=(e_\text{rms}^2 + \sigma_e^2)^{-1}$, where $\sigma_e^2$ is the mean square of galaxy ellipticity in each component, estimated from all galaxies with similar color, brightness, and seeing, and $\sigma_e^2$ is the measurement error \citep{Mandelbaum_2018b}. We further smooth our shear map with a Gaussian smoothing kernel of 2 arcmin in size to reduce the effect caused by the non-uniform weight map (galaxy counts per pixel). For Gaussian statistics, an uneven weight map can be analytically incorporated in the definition of the statistics. However, for non-Gaussian statistics except moment based ones, there is no general method to analytically incorporate the weight map, therefore a proper smoothing is usually needed. We note that compared to our two companion papers \citep{Thiele_2023, Marques_2024}, the smoothing algorithm used in this paper incorporates the weight map in a way closer to the optimal weighting in power spectrum analysis and thus provides lower noise level. The algorithm of smoothing and more discussions are presented in appendix~\ref{app:smoothing}. 
After smoothing, pixels with galaxy density lower than half of the mean density are masked and set to zero. For visualization purpose, we in-paint the masked regions \citep{Pires_2009} when generating Figure~\ref{fig:maps}, but keep the masked regions as zero for statistical analysis. The shear maps are then converted into the mass maps using their arithmetic relation in Fourier space \citep{Kaiser_1993} and implemented assuming periodic boundary conditions for each field of view.

\subsection{Scattering transform and power spectra}

We consider two types of summary statistics of the convergence maps: the power spectrum extracting Gaussian information, and scattering transform coefficients providing substantial non-Gaussian information. 

The scattering transform was recently introduced to cosmology and has desirable properties when performing statistical analyses: it compresses the information in a robust way \citep{Cheng_2020}. It achieves the constraining power of a deep convolutional neural network without the need to train millions of neurons. The number of coefficients used is only comparable to that of the power spectrum, which is smaller than most other non-Gaussian statistics. In practice, such an internal compression of information significantly reduces the estimation noise of the covariance matrix and likelihood emulator. Compared to high-order moment based statistics, the scattering coefficients do not amplify outliers, making the estimator stable under cosmic variance and robust to outliers in real observations.

The scattering transform also serves as a bridge between traditional moment-based statistics and convolutional neural networks. In the context of weak lensing maps, it can measure the scale-dependent sparsity of structures in the field, which reflects how non-linearly the fluctuations have evolved and therefore significantly improve the constraint of $\Omega_\text{m}$.
For a more intuitive understanding, we refer the reader to our pedagogical review \citep{Cheng_2021guide}.

\begin{figure}
    \centering
    \includegraphics[width=0.9\columnwidth]{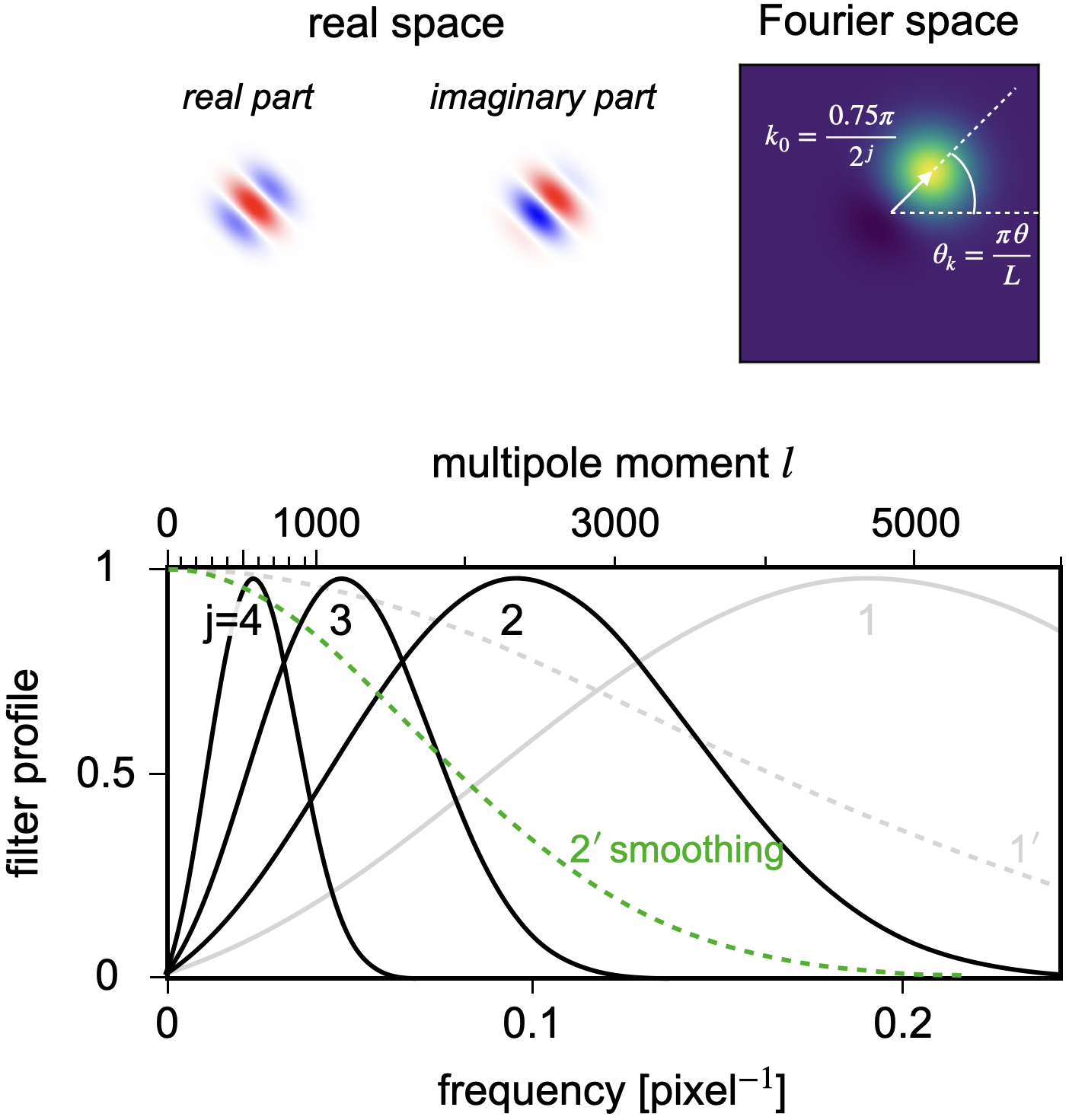}
\caption{\textit{Upper}: The 2D profile of a Morlet wavelet used in the scattering transform. In real space it is a local wave packet, and in Fourier space it is a band pass window. Other wavelets in the same family can be created by dilation and rotation. \textit{Lower}: The radial profile of a family of Morlet wavelets in Fourier space. The pixel size is 0.88 arcmin. In our main analysis, we use maps with 2 arcmin smoothing and wavelets with $j$=2--4.}
    \label{fig:scales}
\end{figure}

From a lensing convergence field $\kappa$, the scattering transform first separates scales by wavelet convolutions, and then applies a point-wise modulus as non-linearity:
\begin{align}
    I_1^{j,\theta} (x,y) = |\kappa * \psi^{j,\theta}|\,,
\end{align}
where '$*$' represents convolution, and wavelets $\psi$ are band-pass filters labeled by their size $j$ and orientation $\theta$. As shown in Figure~\ref{fig:scales}, we use wavelets that are one-sided in Fourier space  and thus complex-valued in real space. The modulus therefore converts fluctuations into their local amplitude (i.e. envelope) so that the $I_1$ fields are local intensity maps of fluctuations in a frequency range. 
In particular we use directional Morlet wavelets, whose profiles in frequency space are close to off-centered Gaussians. The mathematical form is given in Appendix~\ref{app:morlet}.

The 1st-order scattering coefficients are the average of these $I_1$ fields over space and orientations
\begin{align}
    s_1(j) = \langle I_1\rangle_{x,y,\theta}\,.
\end{align}
The $s_1$ coefficients are qualitatively similar to the power spectrum at the central frequency of the wavelet $\psi^j$ -- they just differ in the sense of being $L_1$-norm of the $I_1$ fields instead of the square of $L_2$-norms.

To characterize more non-Gaussian features, each of the $I_1$ fields are further analyzed following the same procedure,
\begin{align}
    I_2^{j_1, \theta_1, j_2, \theta_2}(x,y) &= |I_1^{j_1,\theta_1} * \psi^{j_2, \theta_2}| = ||\kappa * \psi^{j_1, \theta_1}| * \psi^{j_2, \theta_2}|\,\\
    s_2(j_1,j_2) &= \langle I_2\rangle_{x,y,\theta_1,\theta_2}
\end{align}
For the second-order scattering coefficients, as pointed out in \citep{Cheng_2020}, most of the cosmological information lies in the orientation-average coefficients. Therefore, we average over both orientations $\theta_1$ and $\theta_2$. It has also been shown that only coefficients with $j_1 < j_2$ are informative. Intuitively, this is because the modulus converts fluctuations into their local amplitude (envelope), which tends to be smoother than the original fluctuations. The number of scattering coefficients is therefore $J+J(J-1)/2$, where $J$ is the number of wavelet scales used.

Now we consider the power spectrum which carries only the Gaussian information. To cover exactly the same frequency range and make a fair comparison between the Gaussian and non-Gaussian information, we use the same wavelets $\psi$ to weigh and bin the power spectrum, thereby obtaining a wavelet power spectrum,
\begin{align}
    C_l^{\text{wavelet}}(j) = \langle |\kappa * \psi^{j,\theta}|^2 \rangle_{x,y,\theta} = \langle I_1^2 \rangle_{x,y,\theta}
\end{align}
The wavelet power spectrum $C_l^{\text{wavelet}}(j)$ is the averaged power spectrum $C_l$ within the frequency coverage of the wavelets\footnote{Strictly speaking, the averaging kernel is $\langle \int(\Tilde{\psi}^{j,\theta}(k,\theta_k))^2 k d\theta_k \rangle_\theta$, while Figure~\ref{fig:scales} shows $\Tilde{\psi}^{j,\theta}(k,\theta_k=0)$. More discussion can be found in the appendix C of \citet{Cheng_2021guide}.}. For brevity, we shall use $C_l$ to represent $C_l^\text{wavelet}(j)$ in figures and tables and specify the $j$-range in the captions. 
Compared to the normal binning of power spectrum, the wavelet binning guarantees that the scale coverage of the power spectrum is exactly the same as the scattering transform. We also compute the normally binned power spectrum, where 8 bins are set in the range $300<l<3500$ with equal spacing in logarithmic of $l$. They are computed directly from the $\kappa$ map without further treatment on masks. As we follow a forward modeling approach and compute the statistics for observational and mock data in the same way, the mask should not bias the inference.  
When computing the scattering transform and power spectra, we adopt the flat sky approximation and assume periodic boundary condition.
To reduce the influence of masks, when calculating the scattering transform $s_1, s_2$ and wavelet power spectrum $C_l(j)$, we exclude the masked regions in the last step of spatial average, i.e., $\langle \cdot \rangle_{x,y}$ becomes a weighted average with the mask as its weight.

\begin{figure}
    \centering
    \includegraphics[width=\columnwidth]{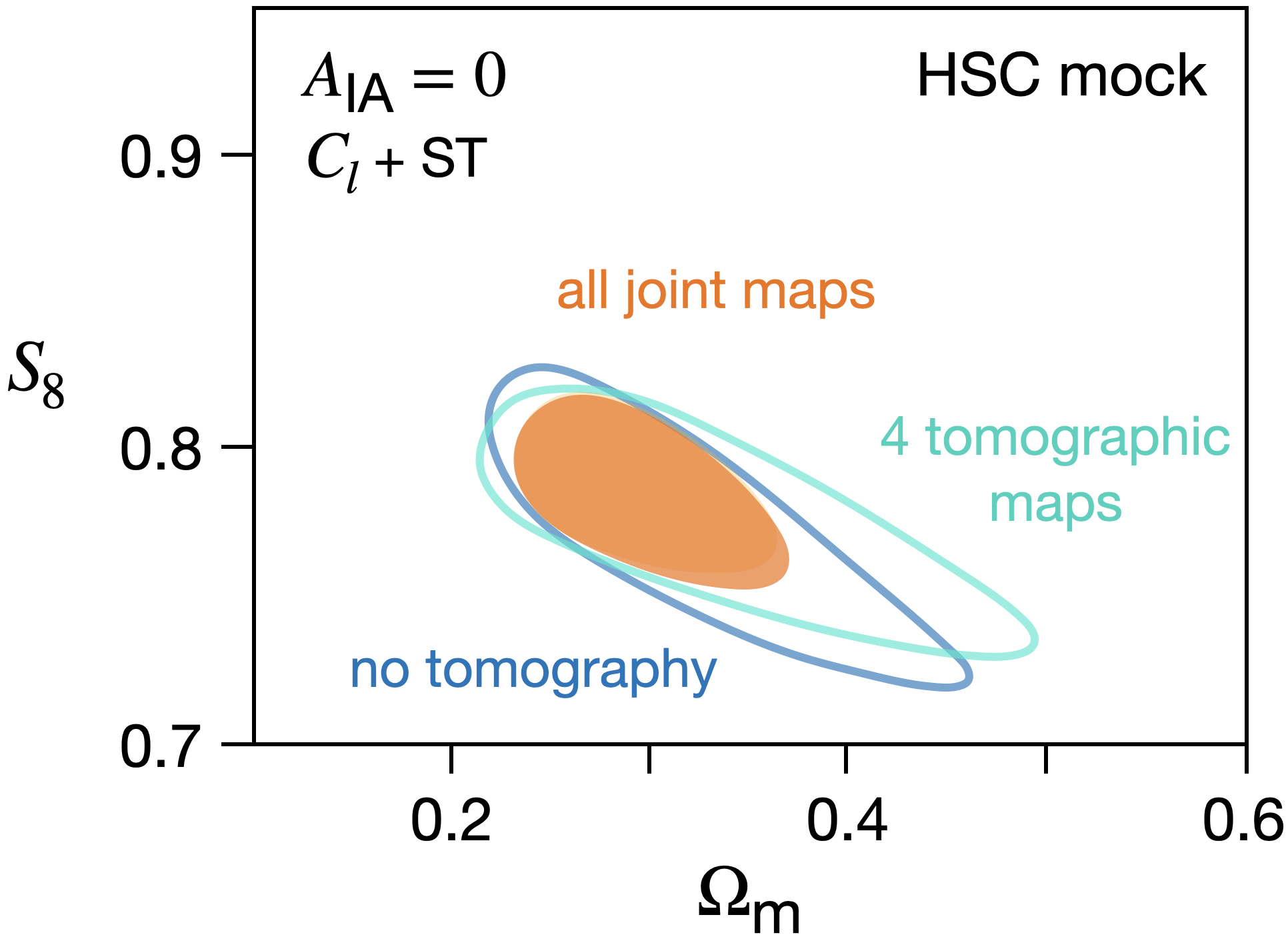}
    \caption{Significant improvement of constraining power by including joint non-Gaussian tomography, shown with mock inference. For simplicity we assume no intrinsic alignment.
    The orange region represents our default setting with 15 joint $\kappa$ maps for scattering transform and 10 maps for the wavelet power spectra (see section~\ref{sec:tomo} for more explanations) with $j$=2--4. It is the 68\% contour of posterior inferred from the mean data vector at fiducial cosmology. The dark blue contour represents the constraint with one map using galaxies with all redshifts, and the cyan contour represents the constraint with four tomographic maps for galaxies from each individual redshift bin.
    }
    \label{fig:mock_tomo}
\end{figure}

\begin{table*}
    \begin{tabular}{ccccccccc}
    \hline
    \text{statistics} & remarks & scale range & \# of coefficients & $\Omega_\text{m}\times1000$ & $\sigma(\Omega_\text{m})\times1000$ & $S_8\times1000$ & $A_\text{IA}$ & $p$-value\\ 
    \hline
    $\boldsymbol{C_l}$\textbf{ + ST} && $j$=2--4 & 120 & \textbf{287} $^{+42}_{-30}$ & \textbf{36} & \textbf{832} $\pm$ 18 & \textbf{0.96} $\pm$ 0.41 & \textbf{0.70} \\
    ST    && $j$=2--4      & 90 & 318 $^{+62}_{-41}$ & 52 & 821 $^{+19}_{-22}$ & $1.14\pm0.40$ & 0.46\\
    $C_l$ && $j$=2--4      & 30 & 300 $^{+83}_{-70}$ & 77 & 820 $^{+18}_{-22}$ & $1.27\pm0.47$ & 0.42\\
    $C_l$ && $l$=300--3500 & 80 & 357 $^{+126}_{-86}$ & 106 & 814 $^{+22}_{-30}$ & $1.21\pm0.48$ & 0.92 \\
    $C_l$ && $l$=300--1900 & 60 & 290 $^{+152}_{-83}$ & 118 & 814 $^{+21}_{-29}$ & $1.00\pm0.55$ & 0.83\\
    \hline
    & no IA &  & & 252 $^{+25}_{-21}$ & 23 & 831 $\pm$ 17 & 0 & 0.43 \\
    &1' smoothing &  &  & 275 $^{+38}_{-27}$ & 33 & 835 $\pm$ 16 & $0.76\pm0.40$ & 0.82 \\
    &TNG baryon   &  &  & 301 $^{+48}_{-33}$ & 41 & 836 $\pm$ 19 & $0.95\pm0.41$ & 0.69 \\
    $C_l$ + ST &no psf & $j$=2--4 & 120 & 299 $^{+47}_{-32}$ & 39 & 834 $\pm$ 18 & 0.92 $\pm$ 0.41 & 0.69 \\
     & \texttt{mizuki} photo-z &  &  & 294 $^{+48}_{-34}$ & 41 & 825 $\pm$ 19 & 1.26 $\pm$ 0.41 & 0.59 \\
     & \texttt{frankenz} photo-z &  &  & 262 $^{+70}_{-42}$ & 56 & 829 $\pm$ 18 & 0.82 $\pm$ 0.33 & 0.57 \\
     & no bin 3 &  &  & 370 $^{+120}_{-79}$ & 100 & 786 $^{+29}_{-36}$ & 0.82 $\pm$ 0.36 & 0.53 \\
    \hline
\end{tabular}
\caption{Summary of our cosmological constraints (marginal medians and 16 and 84 percentiles) using different summary statistics, and the p-values at the best fit of the posterior assuming Gaussian sampling distribution. The default setting (first line with boldface) uses 2 arcmin smoothing, \texttt{MLZ} photo-z and no baryonic effect.} 
\label{tab:obs}
\end{table*}

\subsection{Mock data and emulator}

We follow a simulation-based inference approach to infer cosmological parameters from the observational data, as there is no analytical predictions for the scattering transform statistics so far. To build an emulator for the summary statistics, we use two sets of $N$-body simulations \citep{Takahashi_2017, Shirasaki_2021} to generate multiple realizations of mock shear catalogs customized for our observational data (HSC Y1), and then make mass maps following the same pipeline as on real data. 

Detailed description of the procedures to make the mock catalogs, the cosmological inference framework, and inclusion of systematics are presented in Appendix~\ref{app:mock}, and details about the $N$-body simulations are presented in Appendix~\ref{app:sim}. Below we only briefly summarize the information. 

In our mock dataset, the $N$-body simulations assume the $\Lambda$CDM model. One set is used to sample cosmic variance at a fiducial cosmology \citep{Takahashi_2017}, and the other is used to explore the cosmological dependence of $\Omega_\text{m}$ and $S_8$ \citep{Shirasaki_2021}, which samples 100 cosmology in the range of 0.1 $< \Omega_\text{m} <$ 0.7 and 0.23 $< S_8 <$ 1.1 . After the mock mass maps are made, we build a likelihood emulator of the summary statistics to perform cosmological inference. We fit the simulations with Gaussian sampling distribution whose mean vector changes with $\Omega_\text{m}$ and $S_8$, but covariance matrix is independent of cosmological parameters.

We also include the intrinsic alignment strength $A_\text{IA}$ as a nuisance parameter to be inferred, and we marginalize over two systematic effects: multiplicative bias and PSF residual. For baryonic feedback and different photometric redshift, we only test their influence on different prescriptions but not marginalizing them. We do not include source clustering in our mock data as it is not expected to bias our statistics, which are even functions of the field values \citep[e.g.,][]{Gatti_2024source_clustering}.

\subsection{Non-Gaussian tomography}
\label{sec:tomo}

Correlated Gaussian information between different tomographic maps can be exhausted by computing all the 4 auto- and 6 cross-power spectra. However, when considering non-Gaussian statistics, because many non-Gaussian statistics are defined on single maps, the analysis has usually been performed on individual tomographic maps, and there has not been an agreed way to extract correlated non-Gaussian information. In the context of peak count statistics, \citet{Martinet_2021} proposed to include the cross information by simply computing non-Gaussian statistics on joint maps,
\begin{align}
    \kappa_\text{joint} &= \sum_i N_i\kappa_i
\end{align}
where $N$ is the effective galaxy number density for a tomographic map and the summation is over subsets of ${1,2,3,4}$. There are 11 such combinations plus the 4 individual maps. 
We point out that this is also similar to the way neural networks deal with multi-channel images, and it can be interesting to consider not only the summation, but more general linear combinations, and further look for the optimal choices. Nevertheless, in this study we just follow the idea of \citet{Martinet_2021} and compute the scattering transform on the 15 maps. 
We find that including the cross-map information significantly improves the constraining power 
on $\Omega_\text{m}$ and $A_\text{IA}$. Figure~\ref{fig:mock_tomo} shows the mock inference for $\Omega_\text{m}$ and $S_8$ with no intrinsic alignment. Compared to the usual strategy of using only 4 tomographic maps or only 1 map with galaxies of all redshifts, a factor of two is earned by including the joint maps. In fact, when assuming $A_\text{IA}=0$, it suffices to add only 3 joint maps, $\kappa_{1+2+3+4}$, $\kappa_{2+3+4}$, and $\kappa_{3+4}$ to the 4 individual maps $\kappa_1$, $\kappa_2$, $\kappa_3$, and $\kappa_4$, which provides essentially the same constraining power as using all 15 maps. However, when marginalizing over $A_\text{IA}$, including all the 15 maps improves the constraint of $\Omega_\text{m}$ by about 20\%. A quantitative comparison using real HSC Y1 data is shown in Table~\ref{tab:obs}.

\begin{figure}
    \centering
    \includegraphics[width=\columnwidth]{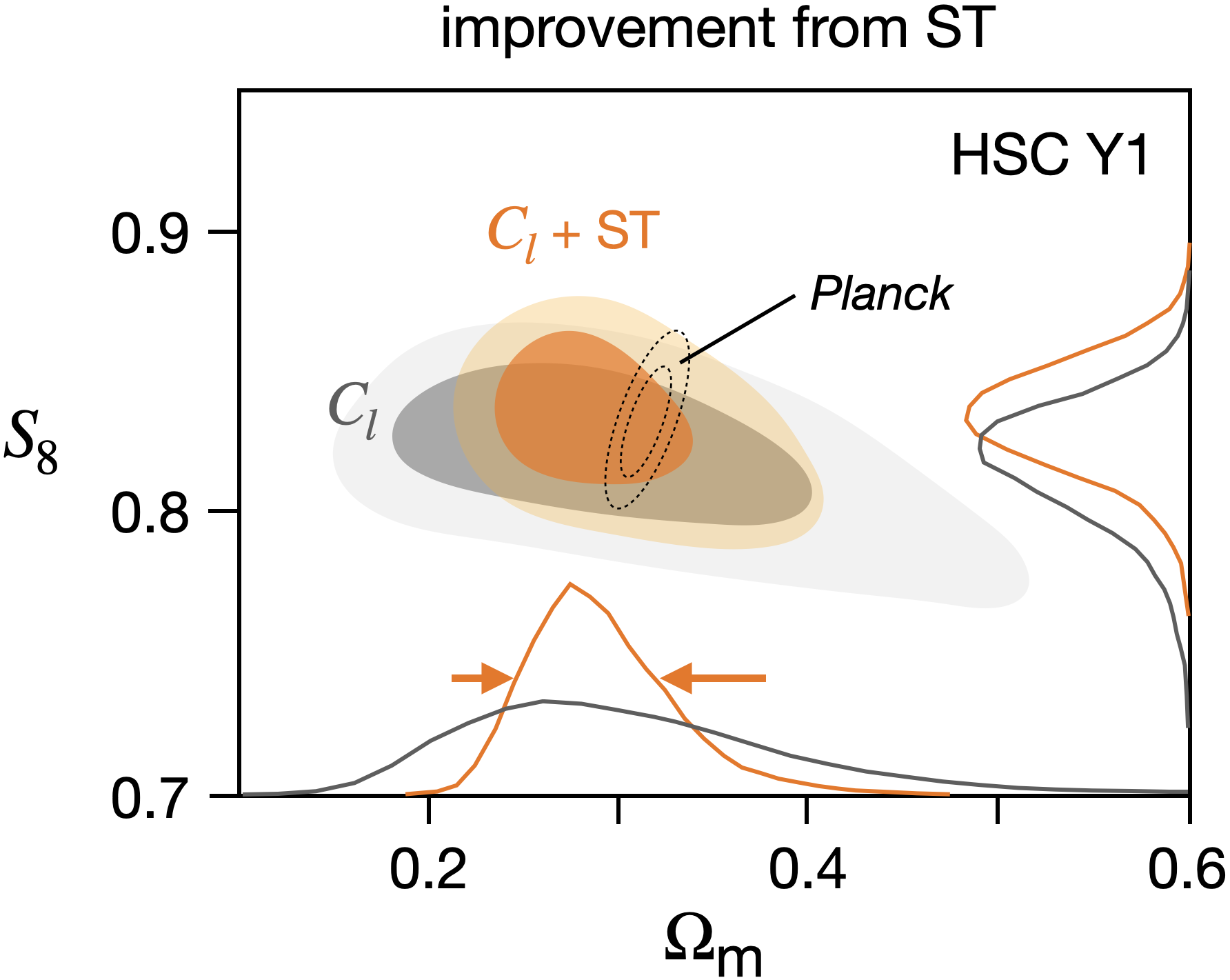}
\caption{Posterior contours (68\% and 95\%) of $\Omega_\text{m}$ and $S_8=\sigma_8 (\Omega_\text{m}/0.3)^{0.5}$ with HSC Year 1 data, using the tomographic scattering transform (ST) and wavelet power spectra covering the same frequency range ($j$=2--4), with intrinsic alignment parameter $A_\text{IA}$ marginalized. The maps are smoothed by a 2' Gaussian kernel. The values of marginal constraints are listed in Table~\ref{tab:obs}. The scattering transform tightens the constraints of $\Omega_\text{m}$ by a factor of 2. 
}
    \label{fig:obs_main}
\end{figure}

\section{Results and discussions}

Figure~\ref{fig:obs_main} shows the cosmological constraints on HSC Y1 data. 
As the default setting, we use maps smoothed by a 2 arcmin Gaussian kernel and wavelets with $j= 2,3,4$ corresponding to central frequencies at $l$ = 600, 1200, 2400 and together covering $l\approx$ 300--3500 (see Figure~\ref{fig:scales}).  In total 90 scattering coefficients (45 $s_1$ and 45 $s_2$) are computed from 15 tomographic maps and joint maps. The wavelet power spectra is computed using the same wavelets for frequency binning and on 10 tomographic and joint maps, which is enough to cover all Gaussian information.

The posteriors of $\Omega_\text{m}$ and $S_8$ are obtained with the intrinsic alignment parameter $A_\text{IA}$ and other nuisance parameters marginalized.  The constraints are also listed in Table~\ref{tab:obs}. Our tight constraint obtained from the scattering transform and power spectra, $\Omega_\text{m}=0.287_{-0.030}^{+0.042}$ and $S_8\equiv \sigma_8(\Omega_\text{m}/0.3)^{0.5}=0.832\pm0.018$, is consistent with the value inferred by \citet{Planck_2020} (TT,TE,EE+lowE+lensing) in both parameters. The small error bar of $\Omega_\text{m}$ and its consistency with \textit{Planck} demonstrate the power and reliability of the scattering transform statistics.

In Table~\ref{tab:obs}, we also show the constraints with several adjustments to the default setting to demonstrate the robustness of our result. In particular, we find that an Illustris-TNG baryonic feedback only shifts the $S_8$ inference up by 0.004 (0.2$\sigma$). Further study of the impact of baryons can be found in \citet{Baryons24}.  Different photo-z codes or PSF-residual modeling do not shift $S_8$ significantly, either.

\subsection{Non-Gaussian estimate improvements}

By adding the scattering transform coefficients to the wavelet power spectra, the constraint on $\Omega_\text{m}$ is improved by a factor of 2\footnote{When compared to the power spectra with 6 bins between $l$ = 300--1900, as adopted by \citet{Hikage_2019}, the improvement is a factor of 3, but apparently, the additional factor of 1.5 comes from extending the frequency range rather than including non-Gaussian information.}.
\citet{Cheng_2020} have shown that on a single-redshift convergence map, the constraining power of the scattering transform is on par with that of the convolutional neural networks reported in \citet{Ribli_2019}. Here we confirm the same constraining power on tomographic analysis of real data by comparing to \citet{Lu_2023}'s convolutional neural net (CNN) results with the same data set.

Recently, new types of neural-net-based models have also been proposed for cosmological inference \citep{Dai_2022, Dai_2024}.  In particular, the multiscale flow model combines the wavelet decomposition idea as used in the scattering transform with the normalizing flow idea to model the likelihood of the whole field, and has been reported to achieve better constraining power than CNN and the scattering transform on simulated weak lensing data \citep{Dai_2024}.  However, its performance in real data with source clustering, survey mask, and other systematics remains to be confirmed.
%

As for the constraining power of $S_8$, we find on both mocks and real data no significant improvement from including non-Gaussian information, consistent with previous forecast on mock data \citep{Cheng_2020}. There has been some confusion in the literature about the tightening of $S_8$ constraint from non-Gaussian statistics. We find that almost all the reported tightening of $S_8$ in the literature can be explained by two reasons. One is the aforementioned mismatch of scales between the non-Gaussian and Gaussian statistics. Many non-Gaussian statistics rely on map smoothing to cut scales, but a Gaussian smoothing kernel does not draw a sharp frequency cut (Figure~\ref{fig:scales}). The second reason is leakage from the tightening of $\Omega_\text{m}$ due to remaining correlation between $S_8$ and $\Omega_\text{m}$, which can be eliminated by optimizing the power index $a$ in the definition of $S_8\equiv \sigma_8 (\Omega_\text{m}/0.3)^a$ or by fixing (conditioning on) $\Omega_\text{m}$. Therefore, we conclude that there is no evidence that non-Gaussian statistics can significantly tighten the error bar of $S_8$. Nevertheless, non-Gaussian statistics do provide consistency checks for the $S_8$ constraint.

\begin{figure}
    \centering
    \includegraphics[width=\columnwidth]{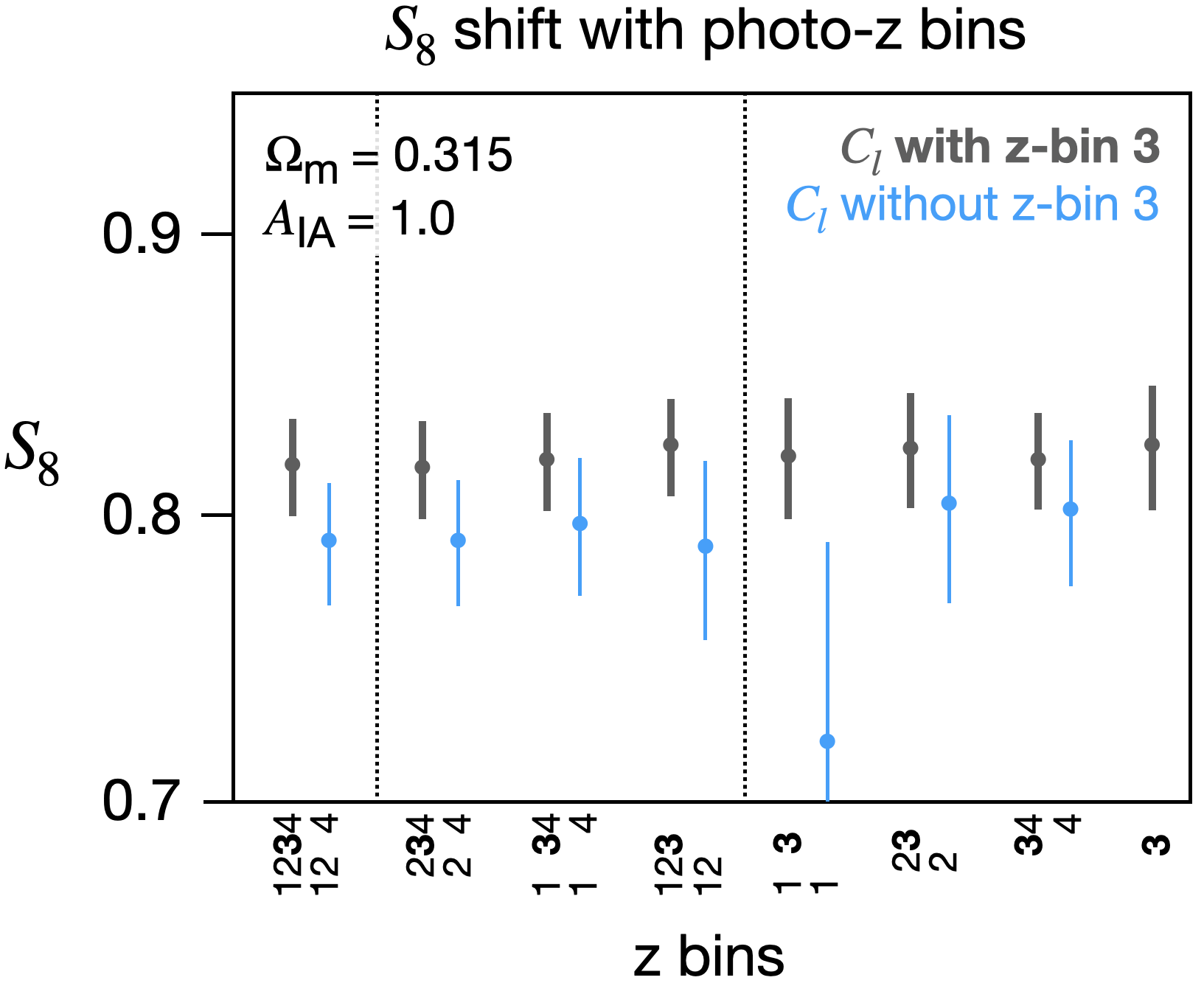}
\caption{Shift of $S_8$ inference when using different subsets of redshift bins. The constraints are performed using tomographic wavelet power spectra (including the cross-spectra). To avoid error leaked from the poorly constrained $\Omega_\text{m}$ and $A_\text{IA}$, the inference is conditioned on $\Omega_\text{m}$=0.315 and $A_\text{IA}$=0. As long as the 3rd redshift bin (photo-z = [0.9-1.2]) is included, the $S_8$ value is stable around 0.82 (gray), otherwise it is around 0.79 (blue), indicating a potential underestimate of redshift in the 3rd bin.
}
    \label{fig:zbin}
\end{figure}

\begin{figure*}
    \centering
    \includegraphics[width=\textwidth]{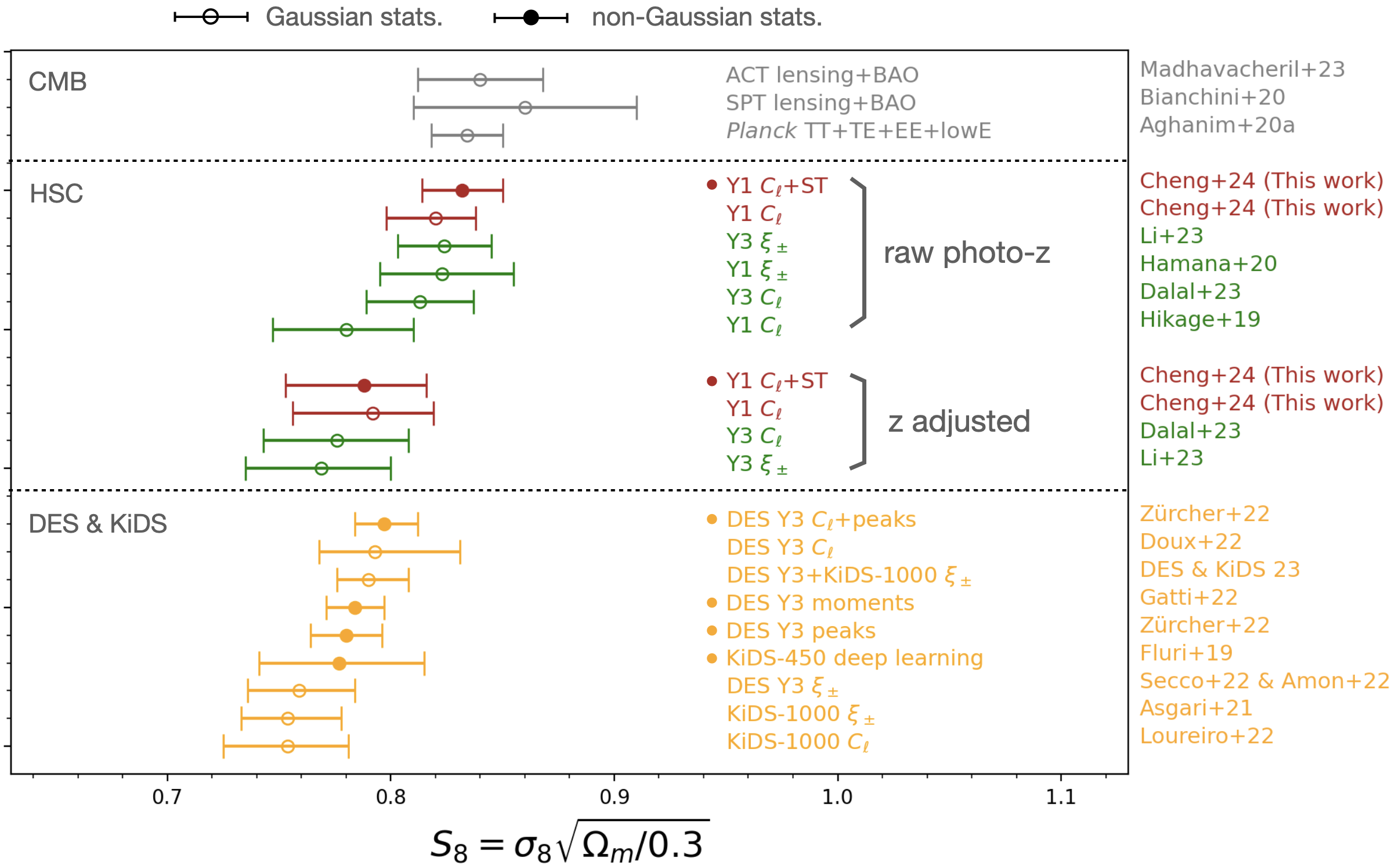}
\caption{Comparison of $S_8$ inference. The upper part shows results from HSC, demonstrating the consistency when using the same raw photometric redshift and the shift towards lower values when adjusting the redshift: for our results, we removed bin 3 from the analysis; for HSC Y3, the authors set free biases of redshift calibration as nuisance parameters and joint constrain them by data. The lower part shows results from other stage-III weak lensing survey and CMB data: \citep{madhavacheril2023atacama, bianchini2020constraints, aghanim2020planck, zurcher2022dark, doux2022dark, kids_des_2023, Gatti_2022, Fluri_2019, Secco_2022, amon2022dark, Asgari_2021, Loureiro_2022}.}
    \label{fig:S8_comparison}
\end{figure*}

\subsection{Influence of photometric redshift on $S_8$}
\label{sec:photoz}

In this section we first point out an internal tension of $S_8$ estimates between different redshift bins, suggesting that photometric redshift is a main limiting factor for an accurate estimate of $S_8$. We then show that when adopting a similar treatment of photometric redshift, our result is consistent with other major results of HSC data analyses.  However, changing the treatment of photometric redshift will substantially change the inferred value of $S_8$.

Figure~\ref{fig:zbin} shows an internal consistency check of photo-z calibration using the inference from different subsets of tomographic maps. For example, the label `12 4' means that all the 6 auto- and cross-power spectra without bin 3, where bin 1--4 correspond to photo-z ranges of [0.3--0.6], [0.6--0.9], [0.9--1.2], and [1.2--1.5]. A systematic trend is found that as long as the 3rd redshift bin is absent, the inferred $S_8$ value is lower than those with bin 3 by 0.03.  Similar patterns are also found when using the scattering transform or binned power spectra as summary statistics, in addition to the wavelet power spectra inference shown in Figure~\ref{fig:zbin}, so this pattern should not be caused by particular statistics but rather linked to the galaxy catalog or map itself.

Though such a lowering of $S_8$ (1.5$\sigma$) may be caused by statistical fluctuations, it may also be caused by an underestimate of the density of high redshift sources, i.e. a tail in the redshift distribution of the 3rd photo-z bin.  Indeed, there are multiple challenges to obtaining reliable photometric redshift estimates when going to deeper magnitude and higher redshifts \citep[e.g.][]{Salvato_2019}.  It requires modeling not only the spectra of galaxies within the redshift range of interest (0.3--1.5 in our case) but all galaxies, and the modeling error for any other galaxies can cause contamination into the redshift range of interest.  
The sensitivity comes from sharp features, mainly the Balmer and Lyman alpha breaks, in galaxy spectra.  
However, the narrow range of optical bands and photometric error lead to degenerate photo-z solutions. As a result, the underlying true redshift distribution of galaxies selected within our photo-z bins tends to have a long tail towards high redshifts, making it difficult to robustly estimate the redshift distribution of the population.  In addition, galaxies above $z$ = 2 in general have high star formation rates and rich emission lines, making it difficult to build reliable templates for galaxy spectra.

In fact, a similar photo-z tension is also found for the 3rd and 4th redshift bins in the HSC Year 3 (Y3) data with 3 times more galaxies \citep{Dalal_2023, Li_2023}.  As shown in Figure~\ref{fig:S8_comparison}, the inferred $S_8$ values also cluster into two groups: with the raw photo-z used, values of $S_8=0.823^{+0.032}_{-0.028}$, $0.824\pm0.021$, $0.813\pm0.024$ are obtained by the correlation analysis of HSC Y1 data \citep{Hamana_2020},  correlation analysis of HSC Y3 data \citep[fig.~12 of][]{Li_2023}, and pseudo-$C_l$ analysis of HSC Y3 \citep{Dalal_2023}. All of the three are consistent with each other and our result\footnote{When comparing their results to ours, we note that for photo-z prescription, they all use the COSMOS re-weighted redshift distribution from galaxies, while we use the stacked redshift posterior with the \texttt{MLZ} code. According to the tests presented in table 7 of \citet{Hikage_2019} and figure 7 of \citet{Hamana_2020}, the inferred $S_8$ is higher by about 0.02 when switching from their method to ours. Therefore, before any special adjustment on photo-z, most HSC analyses, including our simulation-based inference, provide almost identical high values of $S_8$ at 0.82. The only analysis giving a low $S_8$ without photo-z adjustment is the pseudo-$C_l$ analysis of HSC Y1 data reported in \citet{Hikage_2019} ($S_8$ = 0.78). However, they also obtain an unusually low $\Omega_\text{m}=0.16$ not shared by the other analyses, which casts doubts on the reliability of their analysis.} at a `high value' of $S_8$.

On the other hand, when a photo-z adjustment is performed, a similar lowering of $S_8$ is also found in the other HSC analyses. Instead of just excluding bin 3 as we did, the larger area of Y3 data allows for a simultaneous inference (self calibration) of photo-z errors. \cite{Dalal_2023} and \cite{Li_2023} assume that the shape of true redshift distribution in bin 3 and 4 are the same as that provided by the photo-z code, but an unknown overall offset is to be fitted.  It turns out that the HSC Y3 data favors an uncomfortably large offset of $\Delta z \sim 0.1$ and $0.2$ in the two tomographic bins, which dramatically lowers the inferred $S_8$ value: $S_8=0.769^{+0.031}_{-0.034}$ and $0.776\pm0.032$ are obtained from the correlation \citep{Li_2023} and power spectrum \citep{Dalal_2023} analyses, consistent with the lowering of $S_8$ we found in HSC Y1 data when excluding galaxies with photo-z between 0.9 and 1.2. 
Moreover, this internal tension is also observed with non-Gaussian statistics, including in our scattering transform analysis and the peak count analysis of the HSC Y1 data \citep[fig. 9 of ][]{Marques_2024}, despite the different choices of tomography and scale cut.

\subsection{Comparison to other $S_8$ estimates}

Figure~\ref{fig:S8_comparison} shows the comparison of our $S_8$ inference with others in the literature. Regardless of which statistic is used, our results based on the raw redshift estimate from the HSC pipeline \citep{Tanaka_2018} have a `high' value of $S_8$ consistent with \textit{Planck}. 
As discussed in section \ref{sec:photoz}, if we remove background sources with photometric redshifts in the 0.9--1.2 range (based on their "best point estimate" using the  \texttt{Ephor-AB} code), the $S_8$ estimates become relatively lower and more consistent with those obtained with the KiDS\footnote{\url{https://kids.strw.leidenuniv.nl/}} and DES\footnote{\url{https://www.darkenergysurvey.org/}}
weak lensing surveys. 
Because of this internal redshift tension, it is currently difficult to robustly conclude on which value of $S_8$ is correct. 

For a deep survey like the HSC, a significant fraction of galaxies have relatively high redshifts. Therefore, simply removing them from the analysis causes substantial loss of constraining power.
The self-calibration method adopted by \citet{Dalal_2023} and \citet{Li_2023} is a practical solution but still not ideal, as the uncertainty of the redshift distribution is likely from the tail part which is not well captured by the simple overall shift.  To definitively identify the nature of this internal tension from photo-z estimate, it may be essential to obtain independent redshift estimates, for example using the clustering redshift technique \cite[e.g.,][]{Schneider_2006,Newman_2008,Menard_2013,Schmidt_2013,Chiang_2019}.

The above internal tension of $S_8$ suggests that photo-z calibration is a main limiting factor for $S_8$ inference in HSC and other ground based surveys reaching a similar depth, including the Rubin Observatory LSST\footnote{\url{https://www.lsst.org/}} \citep{LSST_2009}. Before robust photo-z estimates become available, improvement of $S_8$ is unlikely to come from optimizing the choice of spatial statistic or machine learning tools \footnote{Nevertheless, the inference of $\Omega_\text{m}$ using non-Gaussian statistics seems robust to the potential photo-z bias.}. As the photo-z bias is likely caused by the ambiguity of spectral structures such as the Balmer and Lyman alpha breaks, space-based surveys such as \textit{Euclid}\footnote{\url{https://www.cosmos.esa.int/web/euclid}} \citep{Laureijs_2011}, Xuntian space telescope survey\citep{Zhan_2021}, and \textit{Roman} space telescope survey\footnote{\url{https://roman.gsfc.nasa.gov/}} \citep{Spergel_2015} which extend photometry to near infrared may significantly alleviate the photo-z problem and therefore draw conclusion on the $S_8$ tension.


\section{Conclusion}

We present the first application of the scattering transform to weak lensing observations. The scattering transform is a summary statistic that borrows ideas from convolutional neural nets, but requires no training and is compact and interpretable (see \citealt[]{Cheng_2021guide} for more details). To best demonstrate its power with real data, we choose to analyze the HSC Year 1 data \citep{Mandelbaum_2018HSC}, which is the deepest wide-field weak lensing survey data available to date.
From a full tomographic analysis of the HSC Y1 data, we conclude:

\begin{enumerate}
    \item \textit{Spatial statistics:} We find that the scattering transform significantly improves the constraint on $\Omega_\text{m}$. It tightens the error bar by a factor of 2 when compared to Gaussian statistics in the same scale range, reaching the constraining power of convolutional neural nets \citep{Lu_2023}. We also find that to improve $\Omega_\text{m}$ constraint, the non-Gaussian information across tomographic maps cannot be ignored, and it can be captured by measuring non-Gaussian statistics on linear combinations of different tomographic maps.

    \item \textit{Photo-z calibration:} We find that the inference of $S_8$ is more limited by photo-z calibration than choosing spatial statistics. When excluding galaxies with photo-z in [0.9-1.2] (the 3rd bin in HSC) from the analysis, the inferred $S_8$ changes from 0.82 to 0.79 independent of which statistic is used, suggesting an under-estimation of redshift of those galaxies.  A similar shift is also found in the analyses of HSC Y3 data \citep{Li_2023, Dalal_2023}. 
    The $S_8$ estimation is currently limited by photometric redshift estimation. This photo-z problem is likely shared by all ground-based weak lensing surveys reaching a similar depth. Calibration with clustering redshift and infrared photometry from space-based telescopes may help to overcome this problem and provide final answer to the $S_8$ tension.
\end{enumerate}




\section*{Acknowledgements}
We thank Joachim Harnois-D\'eraps for providing the simulated intrinsic alignment maps and Ken Osato for providing the kappa-TNG maps.
We thank Alex Drlica-Wagner, Marco Gatti, Kevin~M.~Huffenberger, Nickolas Kokron, Bhuvnesh Jain, and Masahiro Takada for useful discussions.
S. C. thanks Siyu Yao for her constant inspiration and encouragement.
S. C. acknowledges the support of the Martin A. and Helen Chooljian Member Fund, fund from Zurich Insurance Company and the Fund for Natural Sciences at the Institute for Advanced Study.
This work was supported by JSPS KAKENHI Grants 19K14767 and 20H05861 (to MS) and 23K13095 and 23H00107 (to JL). We acknowledge support from the MEXT Program for Promoting Researches on the Supercomputer Fugaku hp230202 (to JL). 
This manuscript has been authored by Fermi Research Alliance, LLC under Contract No. DE-AC02-07CH11359 with the U.S. Department of Energy, Office of Science, Office of High Energy Physics. We thank the Yukawa Institute for Theoretical Physics at Kyoto University, where part of the discussions of this work took place during YITP-T-22-03 on ``Cosmology with Weak Lensing: Beyond the Two-point Statistics'' and YITP-W-23-02 on ``Future Science with CMB x LSS''. Part of this work was performed at the Aspen Center for Physics, which is supported by National Science Foundation grant PHY-1607611. This research used computing resources at Kavli IPMU. This research used resources at the National Energy Research Scientific Computing Center (NERSC), a U.S. Department of Energy Office of Science User Facility located at Lawrence Berkeley National Laboratory, operated under Contract No. DE-AC02-05CH11231. 

\section*{Data availability}

The derived data generated in this research will be shared on reasonable request to the corresponding author.

\bibliographystyle{mnras}
\bibliography{ST_HSCY1}

\appendix

\section{Smoothing of mass map} \label{app:smoothing}

With smoothing, the weighted average should be extended from within each pixel to galaxies in its neighborhood, therefore the weighting becomes a combination of the weights of individual galaxies, $w_i$ and a smoothing kernel $W(\Delta r)$. Approximating the positions of each galaxy by the center of the pixels where they are located, a smooth $\bar{\phi}$ can be obtained by smoothing individually the numerator maps and the denominator map in eq.~\ref{eq:weight},
\begin{align}
    \phi(x,y)_\text{smooth} \equiv \frac{W(\Delta r) * \sum_\text{in pixel (x,y)} w_i \phi_i}{W(\Delta r) * \sum_\text{in pixel (x,y)} w_i}\,,
    \label{eq:smoothing}
\end{align} 
where `$*$' represents convolution on a grid, and the smoothed fields $\phi(x,y)_\text{smooth}$ for the four quantities $\{\boldsymbol{e}, \mathcal{R}, m_\text{tot}, \boldsymbol{c} \}$ are used to generate the $\hat{\boldsymbol{\gamma}}$ map through eq.~\ref{eq:map_pixel}. We use the Gaussian smoothing kernel with standard deviation $\sigma_r$ = 2 arcmin:
\begin{align}
    W(\Delta r) = \frac{1}{2\pi\sigma_r^2}\exp\bigg(-\frac{\Delta r ^2}{2\sigma_r^2}\bigg)\,.
    \label{eq:kernel}
\end{align}
In the literature, as little variation is expected in the map of $m_\text{tot}$ and $\mathcal{R}$ in eq.~\ref{eq:map_pixel}, they are usually treated as constants over a field of view and redshift bin. We have confirmed that replacing $\bar{m}_\text{tot}$ and $\bar{\mathcal{R}}$ maps by the global weighted averages over all galaxies only results in negligible change in the cosmological inference. Therefore, our map making method is essentially the same as \cite{Oguri_2018, Lu_2023}. 

Below we discuss the advantage of this smoothing method compared to a direct convolution of $\bar{\phi}(x,y)$, as adopted in \citet{Thiele_2023, Marques_2024}.
Compared to eq.~\ref{eq:smoothing} where both the numerator and denominator are convolved by the smoothing kernel, an alternative is to directly convolve the pixelized map by the smoothing kernel,
\begin{align}
    \phi'(x,y)_\text{smooth} \equiv W(\Delta r) * 
 \frac{\sum_\text{in pixel (x,y)} w_i \phi_i}{\sum_\text{in pixel (x,y)} w_i}\,,
\end{align} 
However, such a direct convolution is not an optimal estimator because it weighs pixels equally while the effective number of galaxies in each pixel is different. Thus it increases the noise level of the map and reduces the constraining power on cosmological parameters. This is a main difference between the maps used in \citet{Marques_2024, Thiele_2023} and this paper.
To give a quantitative comparison, on the non-tomographic mass map at fiducial cosmology smoothed by a 1 arcmin kernel, our weighted smoothing method provides a correlation coefficient of 0.38 (an S/N of 0.37) between a noisy mock and the noiseless convergence map, whereas the equal-weight smoothing gives a correlation coefficient of 0.32 (an S/N of 0.22). The error bar of inference increases by 50\% in $\Omega_\text{m}$ and 25\% in $S_8\equiv \sigma_8 (\Omega_\text{m}/0.3)^{0.5}$ on both mock and real data. In power spectrum and correlation function analyses \citep[e.g.,][]{Hikage_2019, Hamana_2020}, although there is no explicit smoothing in the map making step, a treatment of weighting equivalent to eq.~\ref{eq:smoothing} is included in the formalism of computing their statistics.

\section{Morlet wavelet}
\label{app:morlet}

The profile of the Morlet wavelets we use is given by
\begin{align}
    \psi^{j,\theta}(\vec{x})&=\frac{1}{\sigma}\exp\left( -\frac{x^2}{2\sigma^2}\right) [\exp(i\vec{k_0}\cdot \vec{x}) - \beta]\,,\\
    \Tilde{\psi}^{j,\theta}(\vec{k}) &= \exp\left[-\frac{(\vec{k}-\vec{k_0})^2\sigma^2}{2}\right] - \beta \exp(-\frac{k^2\sigma^2}{2})\,,
\end{align}
where $\Tilde{\cdot}$ represents Fourier transform, and \begin{align}
    \sigma=0.8\times 2^j\,,\, |k_0| = 0.75 \pi \times 2^{-j}\,,\, \text{angle}(k_0) = \pi \theta / L
\end{align}
specify the dependence on wavelet size index $j$ and orientation index $\theta$. The offset $\beta=\exp\left(-\frac{k_0^2\sigma^2}{2}\right)$ is introduced to guarantee that the wavelets are strictly band-pass instead of low-pass filters: $\Tilde{\psi}(0)=0$. More explanations of the mathematical form can be found in the appendix B of \citet{Cheng_2021guide}.

As a result, the central frequency of the $j=0$ wavelet kernel is defined to be 3/8 pixel$^{-1}$, and when $j$ increases by 1, the wavelet becomes larger in real space by a factor of 2. In this study we use wavelets with $j$ = 2--4. The $j=2$ wavelet roughly covers $l \approx$ 1200--3500, 
$j=3$ wavelet covers $l \approx$ 600--1800,
and $j=4$ wavelet covers $l \approx$ 300--900. 
In principle the sampling of $j$ does not have to be integers, nevertheless we still sample $j$ at integers as we find no significant improvement of cosmological constraint from finer sampling. As for orientations, we follow previous studies and set the total number of orientations to be $L$ = 4, i.e., sampling every 45 degrees.

\section{Simulation based inference method}
\label{app:mock}

\subsection{Mock catalogs}\label{sec:mock}

The mock catalogs are generated following \citet{Shirasaki_2019} by first randomly rotating the observed galaxies and statistically separating the measurement noise from intrinsic galaxy shape,
\begin{align}
    \boldsymbol{e^\text{int}} &= \text{random rotate}\bigg(\frac{e_\text{rms}}{\sqrt{e_\text{rms}^2 + \sigma_e^2}}\bigg) \boldsymbol{e^\text{obs}}\\
    \boldsymbol{e^\text{mea}} &= \sigma_e \boldsymbol{N}
\end{align}
where $e_\text{rms}$ and $\sigma_e$ are read from the observed galaxy catalog \citep{Mandelbaum_2018b} and $\boldsymbol{N}$ is a random variable independently sampled for each galaxy from a standard Gaussian distribution with zero mean and unity variance 
Then, the mock ellipticity catalog is obtained by applying cosmic shear distortion $\boldsymbol{\delta}^\text{sh}$ to the intrinsic galaxy shapes and then adding mock measurement noise \citep{Miralda-Escude_1991, Bernstein_2002}\footnote{With some algebra, one can rewrite these equations in a compact form, which can be found in the literature, in terms of the complex-valued reduced shear $g \equiv \frac{\gamma}{1-\kappa}$ and complex ellipticity: $e = \frac{e^\text{int} + 2g + g^2 e^{\text{int}*}}{1 + |g|^2 + 2\text{Re}(g e^{\text{int}*})}$, by using the relation $\delta^\text{sh}=\frac{2g}{1+|g|^2}$ and $a=\frac{1+|g|^2}{2}$.}
\begin{align}
    e_1 &= \frac{e_1^\text{int} + \delta^\text{sh}_1 + a \delta^\text{sh}_2 (\delta^\text{sh}_1 e_2^{\text{int}} - \delta^\text{sh}_2 e_1^\text{int})}{1 + \boldsymbol{\delta}^\text{sh} \cdot \boldsymbol{e}^\text{int} } + e_1^\text{mea}\\
    e_2 &= \frac{e_2^\text{int} + \delta^\text{sh}_2 + a \delta^\text{sh}_1 (\delta^\text{sh}_2 e_1^\text{int} - \delta^\text{sh}_1 e_2^\text{int})}{1 + \boldsymbol{\delta}^\text{sh} \cdot \boldsymbol{e}^\text{int} } + e_2^\text{mea}\\
    a &\equiv \frac{1 - \sqrt{1-(\delta^\text{sh})^2}}{(\delta^\text{sh})^2} = \frac{1}{1+\sqrt{1-(\delta^\text{sh})^2}}\\
    \boldsymbol{\delta}^\text{sh} &\equiv \frac{2(1-\kappa)}{(1-\kappa)^2+\gamma^2}\boldsymbol{\gamma}\,,
    \label{eq:delta}
\end{align}
where $a$ is a factor close to 0.5 at the small shear limit, and the shear distortion $\boldsymbol{\delta}^\text{sh}$ is obtained by ray-tracing the $N$-body simulations. We also simulate the effect of multiplicative bias $m_\text{tot}$ by replacing $\boldsymbol{\gamma}$ with $(1+m_\text{tot})\boldsymbol{\gamma}$.

The fiducial simulation set \citep{Takahashi_2017, Shirasaki_2019} includes 108 quasi-independent full-sky realizations at one fiducial cosmology with the $\Lambda$CDM model and WMAP9 parameters \citep{Hinshaw_2013}: $\Omega_{\text{cdm}}$ = 0.233, $\Omega_{\text{b}}$ = 0.046, $\Omega_{\text{m}}$ = $\Omega_{\text{cdm}} + \Omega_{\text{b}}$ = 0.279, $\Omega_\Lambda$ = 0.721, $h$ = 0.7, $\sigma_8$ = 0.82, and $n_s$ = 0.97. The sky footprint of the HSC fields was rotated in 21 angles to take full advantage of the large volume of each simulation, resulting in a total of 2,268 realizations of mock HSC lensing maps. More details are described in Appendix~\ref{app:sim} and the references. This set of mock data at the fiducial cosmology is used to estimate the cosmic variance. 

The cosmic-varying simulation set in \citep{Shirasaki_2021} samples 100 different cosmologies with $0.1<\Omega_\text{m}<0.7$ and $0.23<S_8<1.1$. The other parameters are consistent with the {\it Planck} 2015 results \citep{Planck_2016}: $\Omega_b$ = 0.049, $\Omega_\text{m}+\Omega_\Lambda$ = 1, $h$ = 0.6727, and $n_s$ = 0.9645. We make 50 realizations per cosmology by random arrangement of the simulation boxes and further enlarge the size to 150 by adding different realizations of galaxy noise. 

To obtain the $\kappa$ and $\boldsymbol{\gamma}$ in eq.~\ref{eq:delta} to enable generation of the mock galaxy shape catalog, one needs to assign a redshift to each of the galaxies. We follow the same method detailed in \citep{Shirasaki_2019}. In brief, we use the stacked redshift posterior of galaxies provided by the \texttt{MLZ} code, a photo-z code using a self-organizing map, in each tomographic bin as the assumed redshift distribution.

\subsection{Likelihood emulator}

We parameterize the likelihood function by assuming that at each cosmology, the summary statistics $\vec{s}$ follow a multivariate Gaussian distribution under cosmic variance and galaxy shape noise: 
\begin{equation}   
p(\vec{s}|\theta) = \frac{\exp{[-\frac12 (\vec{s}-\vec{\mu}(\theta))^T {\bf C}_\text{fid}^{-1} (\vec{s}-\vec{\mu}(\theta))]}}{(2\pi |{\bf C}_\text{fid}|)^{\frac{\text{dim}(\vec{s})}{2}}}\,.
\end{equation}
Usually, this assumption is justified by the fact that the statistical estimator is an average over a large field of view, for which the central limit theorem Gaussianizes its distribution. This assumption is particularly well justified for the scattering transform coefficients because the modulus operation avoids amplification of outliers and thus accelerates the Gaussianization \citep{Cheng_2021}.

We estimate the mean $\mu (\theta)$ as a function of cosmological parameters $\theta = (\Omega_m, S_8)$ from the set of cosmic varied simulations, by first calculating the sample mean at each cosmology $\hat{\mu} (\theta_i)$ and then fit it with 6th-order polynomials of $(\log \Omega_m, \log S_8)$.
\begin{align}
    \vec{\mu} (\Omega_\text{m}, S_8) = \sum_{0 \leq n+m \leq 6} \vec{a}_{n,m} (\log \Omega_\text{m})^n (\log S_8)^m\,,
\end{align}
where $n$ and $m$ are natural numbers. The reason to use logarithms is that $\mu(\Omega_\text{m}, S_8)$ or its derivatives may have singularity at $\Omega_\text{m}=0$ or $S_8=0$. As the summary statistics may change quickly when $\Omega_\text{m}$ approaches zero, we use a logarithm to regularize this behavior and make the polynomial regression converge faster. The regression is implemented through least square method for each component of $\vec{a}_{n,m}$. 
We verify the robustness of our emulator by fitting it with 5th-order and 7th-order polynomials and find no significant difference, which means our choice does not under-fit or over-fit. The covariance matrix ${\bf C}_\text{fid}$ in our model is fixed and does not change with cosmological parameters. We estimate it only with the 2,268 fiducial realizations that originate from 108 quasi-independent simulations and a full-sky setting. In addition, we find small differences between the fiducial mocks and the emulator prediction based on the cosmo-varied simulations, which is likely caused by the small and adaptive box sizes used in the cosmo-varied simulations. We calibrate the emulator using the mean value of fiducial mocks, which is described in detail in appendix~\ref{app:sim}.

\subsection{Systematics}

To account for observational and astrophysical systematics, we include the following parameters in our emulator: multiplicative calibration error $\Delta m$, galaxy intrinsic alignment strength $A_\text{IA}$, and baryonic feedback strength $A_\text{tng}$. We further model their influence to the summary statistics  with a simple form,
\begin{align}
    \vec{\mu}(\Omega_\text{m}, S_8, A_\text{IA}, \Delta m, A_\text{tng}) =  \vec{b}^{A_\text{IA}} \vec{c}^{A_\text{tng}} \vec{d}^{\Delta m} \vec{\mu}(\Omega_\text{m}, S_8, 0,0,0)\,,
    \label{eq:systematics}
\end{align}
where the coefficients $\vec{b}$, $\vec{c}$, $\vec{d}$ are constants estimated using simulations described below and the operations are meant as element-wise. That parameterization is equivalent to a linear model for the logarithm of our statistics,
\begin{align}
    \log &\vec{\mu}(\Omega_\text{m}, S_8, A_\text{IA}, \Delta m, A_\text{tng}) = \nonumber\\
    &{A_\text{IA}} \log \vec{b} + A_\text{tng} \log \vec{c} + \Delta m \log \vec{d} + \log \vec{\mu}(\Omega_\text{m}, S_8, 0,0,0)\,.
    \label{eq:systematics_log}
\end{align}
We consider it better than a linear model of the original statistics because both the scattering coefficients and the power spectrum are always positive and related to fluctuation strength. This relation is accurate in the neighborhood of the fiducial cosmology, though its accuracy may gradually degrade as one moves away from the fiducial.

\subsubsection{Intrinsic alignment}

To incorporate intrinsic alignment to our emulator for non-Gaussian statistics, we use a mock catalog provide in \citet{Harnois-Deraps_2022} which simulates this effect. This catalog was made following a prescription called $\delta$-NLA, which computes the intrinsic alignment field $\boldsymbol{e}^\text{IA}$ according to the non-linear tidal alignment model (NLA) \citep{Hirata_2004} but also samples galaxies naturally according to the matter density field $\delta$. Therefore, compared to the NLA model, it realistically up-weights regions with higher matter density. The intrinsic alignment strength is still parametrized by $A_\text{IA}$ as in NLA. For simplicity we assume no evolution of $A_\text{IA}$ with redshift. 
Previous studies with HSC Y1 data obtains consistent constraints with different error bar sizes, $A_\text{IA}=0.4\pm0.7$ \citep{Hikage_2019} and $A_\text{IA}=0.9\pm0.3$ \citep{Hamana_2020}, with the NLA model.
According to those approximate range of $A_\text{IA}$, we build mocks with $A_\text{IA}$ = 0 and 1.5, and determine the coefficient $\vec{b}$ in eq.~\ref{eq:systematics} by $\vec{b} = \left( \frac{\langle \vec{s}\rangle_{A_\text{IA}=1.5}}{\langle \vec{s} \rangle_{A_\text{IA}=0}}\right) ^ {\frac{1}{1.5}}$\,.
The average is measured from 400 mocks with 10$\times$10 deg$^2$ field of view, generated by adding different galaxy shape noise realizations to the 10 mock catalogs provided in \citet{Harnois-Deraps_2022}.

\subsubsection{Error in multiplicative bias}

To account for the effect of mis-calibration of multiplicative bias, we model its influence on summary statistics and then marginalize over it. A prior of a Gaussian with 0.01 standard deviation is applied, which is the level of calibration residual estimated from image simulations \citep{Mandelbaum_2018b}. We determine the coefficient $c$ in eq.~\ref{eq:systematics} by its value at fiducial cosmology $\vec{d} = \left(\frac{\langle \vec{s} \rangle_{\text{fid}, \Delta m=+0.01}}{\langle \vec{s} \rangle_{\text{fid}, \Delta m=-0.01}}\right)^{\frac{1}{0.02}}$, which are obtained by adding $\Delta m$ = +0.01 and --0.01 to the multiplicative bias of mock galaxies in the 2,268 realizations of fiducial mock catalogs, while keeping all other ingredients including the random seed for galaxy shape noise unchanged. 
We find that the effect of multiplicative bias error basically shifts $S_8$ by the same fraction, while not changing the inference of $\Omega_\text{m}$. Data itself is not able to constrain $\Delta m$ and it is mainly driven by the prior.

\subsubsection{Baryonic feedback}
We use the kappa-TNG mocks \citep{Osato_2021} to test baryonic effects. They are a set of convergence maps made from the hydrodynamic IllustrisTNG simulations \citep{Nelson_2019}, with 10,000 realizations of 5$\times$5 deg$^2$ maps ray-traced from the highest resolution of the largest box TNG300-1 and its corresponding dark-matter-only counterpart TNG300-1-Dark. Those simulations adopts a cosmology consistent with \citet{Planck_2016}.
Since these mocks are not tailored for the HSC Y1 field of view, the map statistics are not defined identically to the emulator or observation due to mask and edge effects. 
The coefficient $\vec{c}$ in eq.~\ref{eq:systematics} is then determined by the ratio between hydro and dark matter only simulations: $\vec{c} = \frac{\langle \vec{s} \rangle_\text{TNG}}{\langle \vec{s} \rangle_\text{TNG-dark}}$.

\subsubsection{Photometric redshift algorithms}

To test the influence of photometric redshift error, we generate 210 new mocks at the fiducial cosmology using photo-z from two other codes, \texttt{frankenz} and \texttt{mizuki} \citep{Tanaka_2018}. Due to the limited number of additional mocks, we are not able to incorporate $\Delta z_\text{phot}$ as a nuisance parameter 
and marginalize over it.  Therefore, we only apply the ratio $\langle \vec{s} \rangle_{\text{fid}, \texttt{frankenz}}/ \langle \vec{s} \rangle_{\text{fid}, \texttt{MLZ}}$ and $\langle \vec{s} \rangle_{\text{fid}, \texttt{mizuki}}/\langle \vec{s} \rangle_{\text{fid}, \texttt{MLZ}}$ to the emulator and examine the shift of cosmological inference.

\subsubsection{PSF residual}

We also include PSF systematics in our model. Errors in the measurement of point spread function (PSF) and the PSF leakage from imperfect deconvolution lead to additive bias of the shear measurement. The residual of this calibration acts like correlated noises added to the shear map. To account for its effect, we simulate this noise in our mock maps by adding a Gaussian random field to the mock convergence maps, with the four redshift bins sharing the same PSF systematics. We set the power spectrum of this noise using Figure 7 of \citet{Lu_2023}, which corresponds to setting $\alpha_\text{psf}=0.03$ and $\beta_\text{psf}=-0.89$ 
as measured from data by \citet{Hamana_2020}. For simplicity we omit the uncertainty of this power spectrum. Assuming no PSF residual typically shifts $S_8$ up by 0.004 and $\Omega_m$ up by 0.01. A more complicated treatment such as a marginalization of the parametrized power spectrum of the PSF systematics is left for future work.

\subsubsection{Source clustering}

We have not modeled the effect of source clustering, but according to an exploration performed by \cite{Gatti_2023st}, we expect negligible biased for the scattering transform and power spectrum.  Source clustering is the fact that background galaxies are not uniformly distributed but clustered with some correlation of the foreground. Therefore, the observed lensing map is sampled in a biased way and have different statistical properties from uniformly sampled ones as used in our simulations.  Nevertheless, source clustering mostly influences statistics that are odd function of the field (i.e., statistics that change their sign when the field is flipped) \citep[e.g.,][]{Gatti_2024source_clustering}.  As a result, the power spectrum and the scattering transform are not much influenced but source clustering, whereas third-order moment based statistics, phase harmonic coefficients that are odd, and machine-learning methods may be significantly biased if the source clustering effect is not properly modeled.

\section{$N$-body Simulations}\label{app:sim}

The fiducial simulation set \citep{Takahashi_2017, Shirasaki_2019} includes 108 quasi-independent full-sky realizations at one fiducial cosmology with the $\Lambda$CDM model and WMAP9 parameters \citep{Hinshaw_2013}: $\Omega_{\text{cdm}}$ = 0.233, $\Omega_{\text{b}}$ = 0.046, $\Omega_{\text{m}}$ = $\Omega_{\text{cdm}} + \Omega_{\text{b}}$ = 0.279, $\Omega_\Lambda$ = 0.721, $h$ = 0.7, $\sigma_8$ = 0.82, and $n_s$ = 0.97. 
In total 6 independent runs, each with 14 simulations with different box sizes (450--6300$h^{-1}$ Mpc), were implemented to keep the angular resolution roughly the same along the line of sight. Each box has 2048$^3$ particles corresponding to an angular particle grid size of 3.4--6.8 arcmin at the initial condition. The particles in each box are also randomly shifted and ray-traced to create 108 quasi-independent full sky maps with HEALPix resolution $N_\text{side}$=8192. It has been tested that for lensing convergence with sources at $z$ = 1, the power spectrum in simulation is consistent with halo-fit results within 5\% till around $l=3000$ (see Fig.~23 in \citet{Takahashi_2017} and eq.~26 in \citet{Shirasaki_2019}).

The cosmic-varying simulation set in \citep{Shirasaki_2021} samples 100 different cosmologies with $0.1<\Omega_\text{m}<0.7$ and $0.23<S_8<1.1$. The other parameters are consistent with the {\it Planck} 2015 results \citep{Planck_2016}: $\Omega_b$ = 0.049, $\Omega_\text{m}+\Omega_\Lambda$ = 1, $h$ = 0.6727, and $n_s$ = 0.9645.  At each cosmology, four $N$-body simulations with different box sizes and resolutions were run and connected along the light cone to cover a 10 $\times$ 10 deg$^2$ field of view. The box sizes are designed to match the opening angle at all redshifts, with typical values at $\Omega_\text{m}$ = 0.3 to be 280--930 Mpc/$h$. 
The projected particle grid at the initial condition in these simulations has a size of 1.3--2.5 arcmin, which is 2.5 times finer than the fiducial set of simulations.
By randomly shifting each simulation box assuming periodic boundary condition, 50 cosmic shear realizations were generated for each cosmology. From stability tests such as using only a subset of the realizations to build the emulator, we find that 50 realizations are not enough to overcome the numerical noise of the cosmological dependence to mean data vector, so we enlarge the mocks to 150 realizations per cosmology by adding different galaxy noise.

We find that the $\mu(\theta)$ predicted by the emulator at the fiducial cosmology has an offset from the sample mean $\langle \vec{s} \rangle$ of the fiducial simulations. For all tomographic maps, our emulator based on cosmo-varied simulations predicts a lower power spectrum. The deviation is highest for low redshift and for large scales. The deviation in cosmic shear power spectrum (noise removed) is about 20\% for the 1st tomographic bin and 10\% for the 4th bin at $l$ = 300. The deficit gradually changes to zero at around $l$ = 2500 and becomes an excess after that. The excess is expected as the cosmo-varied simulations have higher resolution, but the cause of power deficit is unclear. This difference, dominated by the large scale power deficit, results in an offset in parameter inference. When using the mean of fiducial data vector as input for mock inference, we obtain a best fit of $\Omega_\text{m}=0.261$ and $S_8 = 0.808$, 6.5\% lower in $\Omega_\text{m}$ and 2\% higher in $S_8$ than the fiducial values. We suspect that the deficit of power of our cosmo-varied maps based on \cite{Shirasaki_2021} is caused by the adaptive box size of simulations along the light cone, as this is the main difference from another simulation based analysis of HSC year 1 data \citep{Lu_2023}, where no such effect is observed. The adaptive box size results in the fact that a large angular scale always corresponds to roughly the simulation box size, at which the power may be suppressed due to spectral leakage under windowing. For this study, we just correct for this effect by multiplying the emulator prediction of $\vec{\mu}$ by a factor vector $\langle \vec{s} \rangle _\text{fid} / \vec{\mu}(0.279, 0.791)_\text{emulator}$. A full understanding of the deficit is left for future study.



\bsp	
\label{lastpage}
\end{document}